\documentclass[12pt]{article}
\usepackage{amsmath} 
\usepackage{amssymb}
\usepackage{graphicx} 
\usepackage{natbib}
\usepackage{blindtext}
\usepackage{hyperref}
\usepackage{cleveref}
\usepackage{chngcntr} 
\usepackage{xcolor}   
\usepackage{caption}  
\usepackage{array} 
\usepackage{multirow} 
\usepackage{float}
\usepackage{listings}
\usepackage{booktabs} 
\usepackage[margin=1in]{geometry} 
\usepackage{tabularx} 
\usepackage{setspace}
\usepackage{listings}

\counterwithin{table}{section}
\crefname{table}{Table}{Tables}
\counterwithin{figure}{section}
\crefname{figure}{Figure}{Figures}
\counterwithin{equation}{section}
\crefname{equation}{Equation}{Equations}

\hypersetup{
	colorlinks=true, 
	linkcolor=blue, 
	citecolor=blue,  
	filecolor=blue,  
	urlcolor=blue 
}
\newcommand{\figref}[1]{%
	\hyperref[#1]{Figure~\ref*{#1}}%
}
\newcommand{\tabref}[1]{%
	\hyperref[#1]{Table~\ref*{#1}}%
}
\newcommand{\equref}[1]{%
	\hyperref[#1]{Equation~\ref*{#1}}%
}

\title{The impact of extracurricular education on socioeconomic mobility in Japan: an application of causal machine learning}
\author{Yang Qiang}
\date{2025.01.20}

\begin{document}
	
	\maketitle
	%\clearpage
	
	\begin{abstract}
		%about 10 lines of text
		
		This paper explores the socioeconomic impacts of extracurricular education, specifically private tutoring, on social mobility in Japan. Using data from the 2015 National Survey on Social Stratification and Social Mobility (SSM), we employed a causal machine learning approach to evaluate this educational intervention on income, educational attainment, and occupational prestige. Our research suggests that while shadow education holds the potential for positive socioeconomic impacts, its benefits are undermined by the economic disparities among households, resulting in minimal overall improvement. This highlights the complex mechanisms between individual demographics and educational interventions, revealing promising machine learning applications in this field.
	\end{abstract}
	
	Heterogeneous Treatment Effects
	\clearpage
	
	\tableofcontents
	\clearpage
	
	\onehalfspacing
	
	\section{Introduction}
	%comment: revice the title, but with different words
	% (1.3 ~ 2 pages)
	
	% 1. why this topic is important, 
	%        1/3
	Extracurricular education has become increasingly popular in Japan over the decades \citep{Bray2007}. This "shadow education", including private tutoring, cram schools (\textit{Juku}), and correspondence courses, is often perceived as a pathway to academic achievement \citep{Entrich2015}. According to the latest national survey, about 80\% of elementary and secondary students participated in out-of-school educational activities, and more than 30\% engaged in multiple ones  \citep{MEXT2008}. Such behavior has the potential to bridge or widen existing educational gaps, which in turn affects social equality. Therefore, assessing the effect of extracurricular education on social mobility is important for families and policymakers. 	
	
	% 2. what we already know about it(references)
	%        1/3
	The human capital theory identifies education as an investment in skills, which provides opportunities for socioeconomic status (SES) rises. %Signaling theory implies that education conveys some information about individual capability to the labor market, thereby affecting job choices and income. 
	Empirical studies suggest a positive effect of education on social equality \citep{Gregorio&Lee2002, Abdullah.etal.2013}. However, as Boudon's model of educational expansion suggested, widening access to post-compulsory education does not necessarily lead to increased social mobility \citep{Thompson&Simmons2013}, highlighting the critical nature of education timing. As a complement to compulsory schooling, attending extracurricular education could promote the equality of educational outcomes \citep{Buchmann.etal.2010, HU_FAN_DING_2016}. Research specific to Japan has demonstrated that extracurricular education also contributes to better socioeconomic outcomes \citep{Nakazawa2010, Narisawa2023}. Despite these advantages, the equality in fee-based educational opportunities for students from lower-SES backgrounds has consistently declined since the late 1980s \citep{Nakamura2023}, raising concerns about such interventions.
	
	% 3. what we don't know yet, 
	%    (without causal inference?, linear model v.s. ML, ...)
	%    (ML has been used in econometrics, such as... ,
	%	  not for studying in the socioeconomic mobility in Japan  )
	%        1/3
	Although the literature always presents a positive association between SES and student achievement, several gaps remain. First, many existing studies rely on correlation rather than causation \citep{Black2011}. The true influences for social mobility are usually confounded with residential status, neighborhood environment, and access to education opportunities, given true causality obscured. Without discussion of causality, policies to improve social equality cannot be developed and applied. Second, the latest estimation methods, such as machine learning, have been well used in various fields of econometrics, such as policy evaluation \citep{Athey2019}, quantitative finance \citep{Coqueret2020}, labor market \citep{Fukai2021} and employment \citep{Cockx2023} . However, the study of socioeconomic mobility using ML, especially in Japan, has yet to be fully explored. Given that ML has the ability to model and process high-dimensional data through flexibility, it may provide more insights in this field.	
	
	% 4. what contributations I'd like to do
	%    (linked with part 3)
	This study aims to estimate the causal inference of extracurricular education on the intergenerational transmission of socioeconomic status (SES) in Japan. By employing causal machine learning approaches, the author aims to provide a more detailed understanding of the mechanisms of extracurricular education and social mobility.

	\section{Methodology}
	\subsection{Potential Outcomes Causal Model}
	
	The Potential Outcomes Causal Model, also referred to as the Neyman-Rubin Causal Model (RCM), is a key concept in causal inference. This model conceptualizes causal effects as difference between potential outcomes under randomized experiments, offering a robust framework for causal inference in statistical and econometric research. 
	
	\subsubsection{Definition}
	
	In the Potential Outcomes Causal Model, causality refers to the comparison between potential outcomes that could occur under different treatment conditions. Specifically, a causal effect is defined by the difference between these potential outcomes for an individual unit.
	
	Consider a binary treatment \( W \in \{0, 1\} \), denote \( Y_i(1) \) and \( Y_i(0) \) as the outcomes for unit \( i \) under treatment and control conditions, respectively. Let \(Y_i(W=W_i)\) denote the potential outcome for unit \(i\) when receiving treatment \(W_i\). The observed outcome can be defined as: 
	\begin{equation}
		Y_i = W_i Y_i(1) + (1 - W_i) Y_i(0)
		\label{eq1}
	\end{equation}

	\noindent where \(W_i\) is the treatment indicator (1 if treated, 0 if not).  The Individual Treatment Effect (ITE) is the difference between two potential outcomes:
	\begin{equation}
		\tau_{(ITE)}= Y_i(1) - Y_i(0)
		\label{eq2}
	\end{equation}
	
	Since we cannot observe both \( Y_i(1) \) and \( Y_i(0) \) at the same time, this leads to ``the fundamental problem of causal inference'' \citep{Holland1986}, i.e., that the ITE is impossible to calculate for a specific unit $ i $. To solve this problem, researchers have to rely on various assumptions and approaches to infer causal relationships. 
	
	One common approach is randomized controlled trials (RCT). By randomly assign treatment and control groups, RCT ensures the independence of this assignment from the potential outcome. It is considered the gold standard for establishing causal relationships because randomization produces robust and reliable results, enabling unbiased estimation of the average treatment effect (ATE) and the average treatment effect on the treated (ATT).s
	\begin{equation}
		\tau_{(ATE)} = \mathbb{E}[Y(1) - Y(0)]
		\label{eq3}
	\end{equation}
	
	\begin{equation}
		\tau_ {(ATT)}= \mathbb{E}[Y(1) - Y(0) | W=1]
		\label{eq4}
	\end{equation}
	
	In addition, we may also estimate the conditional average treatment effect (CATE), which represents the difference in expected outcomes between treated and control individuals, given the distribution of specific covariates $X$. 
	
	\begin{equation}
		\tau_ {(CATE)}= \mathbb{E}[Y(1) - Y(0) | X=X_i]
		\label{eq5}
	\end{equation}
	
	\noindent The treatment effects are usually not homogeneous across a population. With CATE, it is possible to assess how treatment effects vary according to specific characteristics of an individual or group, such as age, gender, or socioeconomic status, thereby promoting the development of more personalized and effective intervention strategies. 
	
	In circumstances where it is impractical or unethical to conduct randomized experiments, researchers often turn to observational studies. However, these data can be influenced by confounding variables, which may introduce bias into the estimators. To address this issue, various methods such as propensity score and machine learning can be applied.

	\subsubsection{Identification}
	
	Identification provides a framework within which observed data can be interpreted as reflecting causality. Without proper identification, the differences we observe in outcomes might simply be due to confounding variables, selection bias, or other factors, rather than a true causal effect. In other words, we need to show that the relationship we're interested in can be isolated and measured accurately.
	
	A naïve way to estimate the average treatment effect (ATE) under observational study is to calculate the difference between treated and control groups without making further assumptions. By introducing dummy variables to represent treatment effects $W$, OLS can be used to measure causal effects.  Consider the following linear model:
	\begin{equation}
		Y = \beta_0 + \beta_1 W + \epsilon
		\label{eq6}
	\end{equation}
	
	\noindent where $W$ represents a binary dummy variable (1 if treated, 0 if controlled). In this case, the regression coefficient $\beta_1$ quantifies the estimate of ATE. However, this approach yields biased results, because in addition to ATE, it will produce also confounding bias and heterogeneous effect bias: 
	\begin{equation}
		\begin{aligned}
			&  \mathbb{E}[Y \mid W=1] - \mathbb{E}[Y \mid W=0] \\
			= &\mathbb{E}[Y(1) \mid W=1] - \mathbb{E}[Y(0) \mid W=0] \\
			= &\underbrace{\mathbb{E}[Y(1) - Y(0)]}_{\text{ATE}} \\
			&\quad + \underbrace{\left(\mathbb{E}[Y(0) \mid W=1] - \mathbb{E}[Y(0) \mid W=0]\right)}_{\text{Confounding Bias}} \\
			&\quad + \underbrace{\left(1 - \mathbb{E}[W]\right) 
				\left(\mathbb{E}[Y(1) - Y(0) \mid W=1] - \mathbb{E}[Y(1) - Y(0) \mid W=0]\right)}_{\text{Heterogeneous Effect Bias}}
		\end{aligned}
		\label{eq7}
	\end{equation}

	\begin{enumerate}
		\item \(\mathbb{E}[Y(1) - Y(0)]\) represents the ATE, which is the primary quantity of interest in causal inference. 
		
		\item \(\mathbb{E}[Y(0) \mid W=1] - \mathbb{E}[Y(0) \mid W=0]\) represents the confounding bias or selection bias. This term captures the difference in the expected outcome \(Y(0)\) (i.e., the outcome that would be observed if no treatment were applied) between treated and control groups. If the assignment to \(W\) is influenced by certain factors, such as a tutoring program specifically for good students, it introduces a confounding bias. 
		
		\item \((1 - \mathbb{E}[W]) \left( \mathbb{E}[Y(1) - Y(0) \mid W=1] - \mathbb{E}[Y(1) - Y(0) \mid W=0] \right)\) captures bias due to heterogeneity. Consider a tutoring program aimed at improving student academic performance. Heterogeneous treatment effects are observed when tutoring significantly benefits students who initially struggle (\( \mathbb{E}[Y(1) - Y(0) \mid W=1] \)), but has a lesser or negligible impact on those already performing well (\( \mathbb{E}[Y(1) - Y(0) \mid W=0] \)). Additionally, the term \((1 - \mathbb{E}[W])\) quantifies the proportion of the population not receiving treatment, further highlighting the role of heterogeneous treatment effects. 
		
		% This is particularly true if a large proportion of the population remains untreated, as the potential variation is not reflected in the ATE. 
		
	\end{enumerate}
	
	Therefore, three key assumptions are typically employed to determine the causal effect of treatment from observational data:  
	
	\begin{enumerate}
		\item Stable Unit Treatment Value Assumption (SUTVA)
		\item Unconfoundedness (or Ignorability) Assumption
		\item Overlap (or Positivity) Assumption
	\end{enumerate}
	
	These assumptions are crucial, because they attribute the observed differences between the treated and controlled groups to the treatment $W$ itself. In the next section, we will discuss these assumptions in details.
	
	\subsubsection*{Assumption 1: SUTVA} 
	
	\textit{The potential outcomes for any unit do not vary with the treatments assigned to other units, and, for each unit, there are no different forms or versions of each treatment level, which lead to different potential outcomes \citep{Imbens&Robin2015}.} \\
	
	SUTVA is a fundamental concept in causal inference that implies potential outcomes for each unit $i$ are unrelated to the treatment status of other units. This assumption comprises two components: no interference and no hidden variations. 
	
	% formal defination & simple example
	
	\begin{itemize}
		\item \textbf{No Interference}
		The first component states that the potential outcomes for any given unit are not influenced by the treatment assignments of other units. Formally as:
		\begin{equation}
			Y_i(W_1, W_2, \ldots, W_n) = Y_i(W_i)
			\label{eq8}
		\end{equation}
		
		\noindent where the outcome \(Y_i\) for unit \(i\) under treatment \(W_i\) should depend solely on \(W_i\) and not on the treatment assignments \(W_j\) for any \(j \neq i\). As a simple example, the performance of student $A$ should not be affected by whether student $B$ participates in a tutoring program or not. If $A$ and $B$ are friends who help each other with studies, this assumption would be violated.

		\item \textbf{No Hidden Variations of Treatment}
		The second component states that for each unit, there are no different versions of each treatment level. Specifically, if \(W_i\) refers to the treatment given to unit \(i\), then:
		\begin{equation}
			Y_i(W_i) = Y_i(W_i'), \text{if }  W_i = W_i'
			\label{eq9}
		\end{equation}
		
		For example, to estimate the causal effect of a tutoring program, the assignment of tutors should be randomized. If junior and senior tutors were intentionally divided into two groups, these ''identical`` treatments could lead to different potential outcomes.

	\end{itemize}
	
	% not much interference between units, 
	% make it simple and less critical
	% example, then states that the probability of violation of SUTVA in our study is unlikely to happen due to sparsity of data
	
	With SUTVA, researchers can make more confident estimations of causality. However, it is nearly impossible to verify this assumption based on observational data. Consequently, researchers must rely on their field-specific knowledge to argue for the validity of SUTVA. 
	
	In our case, given the sparseness of the respondent population of the 2015 National Survey of Social Stratification and Social Mobility (SSM) data, which consists of 7,817 units, we can infer that the treatment effect in one unit does not affect the outcomes in other units, thus allowing SUTVA to hold in our study.

	% SUTVA combines two critical conditions that allow for the accurate use of observed data to infer potential outcomes. Firstly, given that the sparsity of the surveyed population in the 2015 SSM data comprises 7,817 units, we can reasonably assume that the no interference hypothesis holds. 
	
	% However, the absence of data regarding the amount of time respondents spend participating in extracurricular education and the quality of such education could lead to potential violations of SUTVA. Such missing information raises concerns over hidden variations in treatment and potential interference. 
	
	% When SUTVA is thought to be violated, it is essential to adopt alternative analytical methods or design considerations to account for these complexities. In this context, we consider coarsening the definition of causal benefits, setting it as the intergenerational transmission effect rather than individual achievement level.

	\subsubsection*{Assumption 2: Unconfoundedness} 
	
	The Unconfoundedness assumption, also known as the Conditional Independence or Ignorability, states that conditional on a set of observed covariates, the treatment is randomly assigned and independent of potential outcomes. It can be formally expressed as:
	\begin{equation}
		Y(0), Y(1)  \perp  W \mid X 
		\label{eq10}
	\end{equation}
	
	Imagine students are randomly selected for the tutoring program, but their scores have been recorded. The assumption would be that any differences in scores post-tutoring are due to this program instead of the student's intelligence or hard work.
	
	Here, the Conditional Average Treatment Effect (CATE) can also be identified as: 
	\begin{equation}
		\begin{aligned}
			\tau_ {(CATE)}&= \mathbb{E}[Y(1) - Y(0) \mid X = X_i] \\
			&= \mathbb{E}[Y \mid W = 1, X = X_i] - \mathbb{E}[Y \mid W = 0, X = X_i]
		\end{aligned}
		\label{eq11}
	\end{equation}
	
	The unconfoundedness assumption is critical for identifying causal effects in observational studies. It allows for comparisons between treated and control groups that are not biased by confounding variables. Again, this assumption is also not testable and require prior-knowledge. In empirical research, it is vital to include a rich array of pre-treatment covariates \(X\) to satisfy the unconfoundedness assumption. The more comprehensive the set of observed covariates, the more plausible this assumption is.

	\subsubsection*{Assumption 3: Overlap}
	
	The overlap assumption, also known as Positivity or Common Support, states that for any combination of covariates, the probability of receiving treatment should be greater than zero. That is to say, each student must has a chance to attend some extracurricular education program. Formally, it can be identified as:
	\begin{equation}
		0 < P(W = w \mid X = X_i) < 1,  \quad \forall w, X_i 
		\label{eq12}
	\end{equation}
	
	% http://www.stat.columbia.edu/~gelman/arm/chap10.pdf
	% https://ehsanx.github.io/psw/balance.html
	\noindent One approach to evaluating overlap is to perform a statistical check, of which a popular method is the \textbf{Standardized Mean Difference (SMD)}. For the treated and control groups, let \(\mu_1 = \mathbb{E}[X_i \mid W_i = 1]\) and \(\mu_0 = \mathbb{E}[X_i \mid W_i = 0]\) denote the means; let \(\sigma_1^2 = \mathbb{V}(X_i \mid W_i = 1)\) and \(\sigma_0^2 = \mathbb{V}(X_i \mid W_i = 0)\) denote the variances, respectively. Then SMD can be defined as the normalized mean differences divided by an estimate of the within-group standard deviation, formally as: 
	\begin{equation}
		\Delta_{std} = \frac{|\mu_1 - \mu_0|}{\sqrt{(\sigma_1^2 + \sigma_0^2)/2}},
		\label{eq13}
	\end{equation}
	
	\noindent which allows the comparison of variables at different scales. Although there is no single rule for determining what value suggests a significant imbalance, an SMD less than 0.1 is often used to indicate a negligible difference between treatment groups \citep{Austin2011}.
	
	Overlap allows for a valid and unbiased comparison between treated and control groups. Otherwise, any causal inference would have to rely heavily on model-based assumptions rather than being directly informed by the data. 
	
	\subsubsection*{Summary} 
	
	Potential outcome causal inference provides a rigorous and structured approach to understanding and estimating causal effects from randomized data. With these foundational assumptions, researchers can proceed with confidence to the estimation phase, employing statistical techniques that leverage these conditions to estimate causal effects accurately.

	\subsection{Estimating Treatment Effects} 
	
	In observational studies, the concern of biased ATE estimation remains. As the number of variables increases, the number of possible combinations and interactions between them grows exponentially. This dimensionality problem makes it difficult to identify true causality as well as to distribute the effects of different variables. 
	
	\citet{Rosenbaum1983} propose an equivalent	and feasible estimation strategy based on the concept of	\textbf{Propensity Score}. Bias from the imbalance of the covariate can be corrected by conditioning on the univariate propensity score, not the covariate vector $X_{i}$. Formally, the propensity score is the conditional probability of receiving the treatment given the pre-treatment variables. For a given unit \(i\) with covariates \(X_i\), it is defined as:
	\begin{equation}
		e(X_i) = \Pr(W_i = 1 \mid X_i)
		\label{eq14}
	\end{equation}
	
	The propensity score provides a robust method by which researchers can approximate the conditions of a randomized experiment, enabling more accurate estimation of treatment effects. By using regression and propensity score, the potential outcome can be expressed as following ways: 
	\begin{equation}
		\begin{aligned}
			\mathbb{E}[Y(w) \mid X = x] &= \mathbb{E}[Y \mid W = w, X = x] := m(w, x) \\
			&= \mathbb{E} \left[ \frac{\mathbf{1}[W = w] Y}{e_w(x)} \mid X = x \right] \\
			&= \mathbb{E} \left[ m(w, x) + \frac{\mathbf{1}[W = w] (Y - m(w, x))}{e_w(x)} \mid X = x \right]
		\end{aligned}
		\label{eq15}
	\end{equation}
	
	\noindent where $m(w, x)$ represent the expected outcome under treatment assignment. Therefore, there are three corresponding estimation strategies for ATE: Outcome Regression \citep{Rubin1979}, Inverse Probability Weighting (IPW) \citep{Rosenbaum1987} and Augmented IPW (AIPW)/Doubly Robust \citep{Robins1994}. 
	
	\subsubsection{Outcome Regression}
	
	Outcome regression estimates causal effects by OLS model for outcomes on the treatment and covariates. Unlike randomized trials, observational data may have confounders that bias the estimation of treatment effects. The idea of this approach is to quantify how changes in a treatment variable are associated with changes in the outcome variable while controlling for potential confounders. Assume the treatment effect is correctly specified, ATE can be estimated as: 	
	\begin{equation}
		\begin{aligned}
			\hat{\tau}_{reg} = \frac{1}{n} \sum_{i=1}^n \hat{m}(w, X_i) = \mathbb{E} \left[ \hat{\mu}_1(X_i) - \hat{\mu}_0(X_i) \right]\\
		\end{aligned}
		\label{eq16}
	\end{equation}
	
	\noindent where $\mu_w(X_i) = \mathbb{E}[Y(w) \mid X_i ]$ indicates predicted outcome for individual $i$ under treatment condition, given their covariates $X_i$. The difference $ \hat{\mu}_1(X_i) - \hat{\mu}_0(X_i) $ represents the estimated individual treatment effect, and the average of these differences across all individuals provides an estimate of the ATE.
	
	To illustrate outcome regression, we specify two models for each potential outcome with known treatment assignment distributions:
	
	\begin{equation}
		\begin{aligned}
			\mu_0(X_i) = \mathbb{E}[Y(0) \mid X_i]\\
			\mu_1(X_i) = \mathbb{E}[Y(1) \mid X_i]			
		\end{aligned}
		\label{eq17}
	\end{equation}
	
	\noindent this can be achieved by using linear regression that includes a dummy variable for the treatment \(W_i\):
	\begin{equation}
		Y_i = \beta_0 + \beta_1 X_i + \beta_2 W_i + \epsilon_i
		\label{eq18}
	\end{equation}
	
	\noindent where $\beta_2$ estimates the ATE. By including covariates \(X\), outcome regression adjusts for confounders that could bias the treatment effect estimate, which is essential in non-experimental settings where randomization is not feasible.
	
	However, there are some practical concerns when using outcome regression for causal inference. First, the valid estimation depends heavily on the accuracy of the linear model specification. Mis-specification or unmeasured confounders can lead to biased estimates. Second, highly correlated covariates can make it difficult to estimate individual effects, thus affecting the stability of the model. 
	
	% Outcome regression emphasizes the importance of model specification, careful selection of covariates, and thorough diagnostics to ensure robust and reliable causal estimates. While powerful, it requires caution regarding the assumptions and potential limitations inherent in observational data analysis.
	
	\subsubsection{Inverse Probability Weighting(IPW)}
	
	The Inverse Probability Weighting (IPW) estimator involves weighting individuals by the inverse of their propensity score $e(X_i)$. This adjusts for the fact that some individuals are more likely to receive treatment than others due to their covariate values, thereby creating a pseudo-population where treatment is independent of covariates. 
	
	For binary treatment $W$ and well-estimated propensity scores $e(X_i)$ (using logistic regression or suitable methods): 
	
	Let $W$ indicates the binary treatment assignment, if the propensity score $e(X_i)$ is well estimated using logistic regression or any suitable methods, 
	
	\begin{itemize}
		\item Treated individuals $(W = 1)$ are weighted by \( \frac{1}{e(\mathbf{X})} \)
		\item Control individuals $(W = 0)$ are weighted by \( \frac{1}{1 - e(\mathbf{X})} \)
	\end{itemize}
	
	These weights enable the sample to resemble a randomized experiment, ensuring similar covariate distributions across groups. Therefore, IPW estimator can be expressed as:		
	\begin{equation}
		\hat{\tau}_{IPW} = \frac{1}{n} \sum_{i=1}^n \left( \frac{W_i Y_i}{\hat{e}(X_i)} - \frac{(1-W_i) Y_i}{1 - \hat{e}(X_i)} \right)
		\label{eq19}
	\end{equation}
	
	As a semi-parametric method, IPW doesn't require a specific functional form for the propensity score model. This flexibility allows it to accommodate complex relationships between covariates and treatment assignments. However, the reliability of IPW depends heavily on accurately specifying the propensity score model. Extreme weights, often occurring when propensity scores are near 0 or 1, can increase estimate variance. Therefore, large sample sizes are typically necessary for stable results, particularly with rare treatments or outcomes.
	
	\subsubsection{Augmented IPW (AIPW)/Doubly Robust}
	
	Augmented Inverse Probability Weighting (AIPW), also known as doubly robust estimation, is a powerful technique in causal inference that combines features of both IPW and outcome regression. This approach provides robustness against mis-specification of either model, making it a popular choice for estimating causal effects from observational data.
	
	Based on the given equations for outcome regression \( \hat{\tau}_{\text{reg}}\) and IPW \(\hat{\tau}_{IPW}\), the AIPW estimator can be expressed by combining these approaches:	
	\begin{equation}
		\begin{aligned}
			\hat{\tau}_{AIPW} &= \frac{1}{n} \sum_i^n \left( \hat{m}(w, X_i) + \frac{\mathbf{1}[W_i = w] (Y_i - \hat{m}(w, X_i))}{\hat{e}_w(X_i)} \right) \\
			&= \frac{1} n \sum_{i=1}^n \left[ \hat{\mu}_1(X_i) + \frac{W_i (Y_i-\hat{\mu}_1(X_i))}{\hat{e}(X_i)}\right] -  \frac{1} n \sum_{i=1}^n \left[ \hat{\mu}_0(X_i) + \frac{(1-W_i) (Y_i-\hat{\mu}_0(X_i))}{1-\hat{e}(X_i)} \right]
		\end{aligned}
		\label{eq20}
	\end{equation}
	
	\noindent which allows the AIPW estimator to remain consistent if either the propensity score model or the outcome regression model is correctly specified, a property known as double robustness \citep{Joseph2007}. 
	Besides, the AIPW estimator has other attractive statistical properties. \citet{Chernozhukov2018} show that the AIPW estimator in \equref{eq20} is consistent, asymptotically normal and semi-parametrically efficient if nuisance parameters are high-quality and cross-fitted predictions. This supports the use of machine learning techniques to estimate the nuisance parameters $\hat{m}(w, X)$ and $\hat{e}_w(X)$, enhancing the estimator's reliability.
	
	\subsubsection*{Summary} 
	This section explored three key methods for estimating causal effects in observational studies. Outcome Regression uses covariates to predict potential outcomes, while IPW adjusts for treatment likelihood by weighting individuals according to their propensity scores. AIPW combines the strengths of both methods, offering a doubly robust approach that remains valid if either model is correctly specified. These techniques pave the way for more advanced methodologies like Causal Tree and Causal Forest, providing a framework to enhance causal inference by capturing complex interactions and heterogeneities within data.

	\subsection{Estimating Heterogeneity with ML}
	
	Machine Learning refers to the study of computer algorithms that improve automatically through experience \citep{Mitchell1997}. There has been significant progress in leveraging machine learning techniques into econometrics. \citet{Belloni2014} introduced \textbf{Double Selection} that enhances variable selection in the presence of high-dimensional datasets, ensuring that important predictors are not overlooked while also managing potential biases. \citet{Chernozhukov2018} developed the framework of \textbf{Double ML}, which accurately estimates causal parameters by accounting for complex relationships and potential confounders in the data. Recently, \textbf{Causal Forest} has emerged as a powerful tool in estimating heterogeneous treatment effects (HTE), combining the flexibility of machine learning models with the rigorous statistical properties necessary for causal estimation\citep{Athey2019grf}.

	\subsubsection{Causal Tree}
	
	Causal Tree, as defined by \citet{Athey2016}, is a decision tree framework designed specifically to estimate HTE. It partitions data into subgroups based on covariates, aiming to maximize the differences in treatment effects between subgroups. These trees use recursive partitioning techniques and adapt them to a causal inference setting by optimizing for treatment effect estimation rather than prediction accuracy. 
	
	Specifically, Causal Trees extend the framework of decision trees to estimate conditional average treatment effects (CATE). In a decision tree, the goal is to predict outcomes based on covariates, minimizing prediction error. In contrast, Causal Trees aim to minimize the mean squared error (MSE) of treatment effects across different subgroups. The construction of a Causal Tree is similar to that of traditional decision trees but with important modifications to account for treatment assignment and causal inference. The algorithm can be summarized in the following steps:
	
	\begin{enumerate}
		\item \textbf{Partitioning the Covariate Space:} The algorithm recursively partitions the covariate space \( X \) into subgroups (or leaves) using a greedy algorithm. At each step, it selects the binary split that minimizes the MSE of the treatment effect in the resulting subgroups. 
		
		\item \textbf{Ensuring Sufficient Balance:} A critical step in Causal Tree construction is ensuring that each leaf contains both treated and control units. Without sufficient representation from both groups, it would be impossible to estimate treatment effects. This is achieved by imposing constraints on the minimum number of treated and control units in each leaf.
		
		\item \textbf{Cross-Validation:} Cross-validation includes resampling and sample splitting methods that use different portions of the data to test and train a model on different iterations. This ensures that the tree generalizes well to unseen data by evaluating the performance of different tree depths on left-out data.
		
		\item \textbf{Pruning the Tree:} After the tree is built, the partition is pruned to remove leaves that provide minimal improvement in treatment effect estimation. This process reduces the complexity of the tree and enhances interpretability.
		
		\item \textbf{Honesty in Estimation:} One of the key innovations in Causal Tree is the use of sample splitting, called \textit{honesty}. The data is split into two samples, where one part of the data is used for building the decision tree structure, and the other part is used for estimating treatment effects to prevent overfitting and maintain unbiased estimates. This approach reduces overfitting and ensures unbiased estimations. 
	\end{enumerate}
	
	Causal Trees offer several advantages in causal inference. It provides a clear and interpretable model, where the treatment effects for each subgroup can be easily understood. By incorporating sample splitting, Causal Trees offer unbiased estimates of treatment effects with valid confidence intervals, addressing concerns about overfitting and ensuring robust inference. Besides, the recursive partitioning algorithm allows for the automatic discovery of subgroups in the data where treatment effects differ, providing valuable insights into HTE.

	%	https://hal.science/hal-04177493/document
	While Causal Trees are a powerful tool for estimating heterogeneous treatment effects, they also have limitations. One significant limitation is that they produce discrete treatment effect estimates, meaning that all units within a leaf are assigned the same estimated treatment effect. This can lead to discontinuities in the estimates, especially for units near the boundary between leaves. Furthermore, the estimates may vary depending on how the data is split, potentially reducing the robustness of the method. To address these issues, Causal Forest has been developed.

	\subsubsection{Causal Forest}
	
	Like random forest and regression trees, Causal Forest approach, as implemented in the grf R package\citep{GRF2024}, aggregate multiple trees to provide more stable estimates of CATE. By averaging across many trees, Causal Forest could produce reliable estimates of treatment effects while maintaining interpretability. This flexible, non-parametric approach allows for the identification of HTE without making strong parametric assumptions. 	
	
	At the core of the Causal Forest approach is the partially linear model, which decomposes the observed outcome into components:
	\begin{equation}
		\begin{aligned}
			Y_i = f(X_i) + W_i \tau(X_i) + \epsilon_i
		\end{aligned}
		\label{eq21}
	\end{equation}
	
	This partially linear model allows the treatment effect, \(\tau(X_i)\), to vary as a function of covariates \(X_i\), enabling the identification of heterogeneous effects. To isolate treatment effects, residuals are computed for both outcomes and treatments as:
	\begin{equation}
		\begin{aligned}
			Y_i^{\text{res}} = Y_i - m(X_i), \quad W_i^{\text{res}} = W_i - e(X_i)
		\end{aligned}
		\label{eq22}
	\end{equation}
	
	\begin{itemize}
		\item \( m(X_i) = \mathbb{E}[Y_i \mid X_i] \): The expected outcome given covariates.
		\item \( e(X_i) = \mathbb{E}[W_i \mid X_i] \): The expected treatment probability (propensity score) given covariates.
	\end{itemize}
	
	This adjustment removes systematic variability due to covariates, isolating the causal effect of the treatment which is modeled as a function of the treatment residuals:
	
	\begin{equation}
		\begin{aligned}
			Y_i^{\text{res}} = \tau(X_i) W_i^{\text{res}} + \epsilon_i
		\end{aligned}
		\label{eq23}
	\end{equation}
	
	This formulation allows us to estimate the Conditional Average Treatment Effect (CATE), \(\tau(X_i)\), which quantifies the treatment effect for individuals with covariates \(X_i\).
	
	Finally, the treatment effect \(\tau(X_i)\) can be estimated by optimizing the following objective function:
	\begin{equation}
		\begin{aligned}
			\hat{\tau}^{\text{cf}}(x) = \arg\min_{\tau} 
			\left\{ 
			\sum_{i=1}^{N} \alpha_i(x) \left[ (Y_i - \hat{m}(X_i)) - \tau(x)(W_i - \hat{e}(X_i)) \right]^2 
			\right\}
		\end{aligned}
		\label{eq24}
	\end{equation}
	
	\noindent where weights ${\alpha_i(x)}$ are specific to $x$ and can be selected by cross-validation.
	
	Causal Forest is a flexible, non-parametric method for estimating heterogeneous treatment effects. By averaging over multiple Causal Trees and using adaptive subsampling techniques, Causal Forest can efficiently handle high-dimensional data and provide valid statistical inference. Their ability to estimate smooth treatment effects makes them a valuable tool in policy evaluation and other areas where understanding HTE is crucial.

	\section{Data Description}
	
	\subsection{SSM Survey}
	\subsubsection{Introduction of SSM Survey}

	The national survey of social stratification and social mobility (SSM) is one of the most traditional large-scale social surveys in Japan. Conducted every ten years since 1955, this survey examines key aspects of individuals' socioeconomic and demographic conditions, including income, education, occupation, and family background. The primary objective of the SSM survey is to provide systematic data on the structure of Japanese society, serving as an important resource for academic research and policy-making. 
	
	Our study used data from the 2015 SSM Survey, the seventh in the series. This survey focused on understanding changes and current conditions of social stratification in Japan, with a particular focus on demographic changes such as the rapid aging of the population and declining birthrate. These challenges have significant implications for labor market dynamics, intergenerational mobility, and socioeconomic inequality, making the survey a vital tool for assessing Japan’s societal transformations.
	
	The 2015 SSM survey used a two-stage stratified random sampling method, categorizing municipalities into five strata based on population size. From 805 designated areas, individuals aged 20-79 were systematically sampled using Basic Resident Registers. This multistage sampling approach achieved comprehensive national coverage, with an intended sample size of 16,100. From this, 15,605 were selectively engaged, resulting in 7,817 valid responses, achieving a response rate of 50.1\%. Post-stratification weights were applied correct for non-response and align the sample distribution more closely with the population.

	\subsubsection{Variables Explanation}
	
	The outcome variables are intergenerational transmission of educational attainment (ITE), income (ITI), and occupational prestige (ITOP). They were calculated as the absolute differences between the respondent's standardized scores in education, income, and occupational status and those of their parent(s) \citep{Ryabov2020}. For each intergenerational transmission status \(i\), the outcome is defined as: 
	
	\begin{equation}
		\ IT_i = \left(SS_{ic}-SS_{ip}\right)
		\label{eq25}
	\end{equation}
	
	\noindent where \(IT_{i}\) is intergenerational transmission of \textit{i}-th dimension of social status (\textit{i} = education, income or occupational prestige); \(SS_{ic}\) and  \(SS_{ip}\) represent the standardized scores of the \textit{i}-th dimension of social status for child and parent, respectively. 
	
	To ensure comparability, data were standardized to a unified scale. Educational levels before and after the 1945–1952 reform were specified \citep{Nakano1973}; parental income was estimated using principal components analysis (PCA) applied to 19 consumer durable items \citep{Filmer2001}; occupational status was categorized into the EGP class scheme based on job descriptions \citep{Kanomata2008}. Finally, the outcome measures \(IT_{i}\) were calculated by subtracting the standardized parental scores from the respondent's scores for each dimension, giving comparable and interpretable measures of socioeconomic mobility.

	In the SSM dataset, extracurricular education are represented by cram schools (\textit{Juku}), private tutoring, and correspondence courses. This study selects private tutoring as the treatment variable for its personalized educational experience, making it a clearer indicator of customized intervention. Other pre-treated demographic features were selected as the explanatory variables, such as gender, age cohort, and living prefecture. A complete list of all study variables for the total sample is provided in \tabref{tab1}. 
	
	\begin{table}[hbt!]
		\centering
		\caption{
			\textmd{Variables Explanation.}
		}
		\label{tab1}
		\begin{tabularx}{\textwidth}{@{}p{5cm}*{1}{>{}X}@{}}
			%\begin{tabularx}{@{}p{5cm}p{11cm}@{}}
			\toprule
			\textbf{Variable Name} & \textbf{Description} \\
			\midrule
			\textbf{Outcome Measures} & \\
			\textmd{Intergenerational Transmission of Educational Attainment (ITE)} & Difference in standardized educational attainment scores between respondents and their parents. \\
			\textmd{Intergenerational Transmission of Income (ITI)} & Difference in standardized income scores between respondents and their parents. \\
			\textmd{Intergenerational Transmission of Occupational Prestige (ITOP)} & Difference in standardized occupational prestige between respondents and their parents, based on EGP class scheme. 
			\\
			\textbf{Treatment Variables} & \\
			\textmd{Cram school (\textit{Juku})} & Experience with Cram school (\textit{Juku}) for more than 6 months during elementary or secondary school (1 = yes, 0 = no). \\
			\textmd{Private tutoring} & Experience with private tutoring for more than 6 months during elementary or secondary school (1 = yes, 0 = no). \\
			\textmd{Correspondence courses} & Experience with correspondence courses for more than 6 months during elementary or secondary school (1 = yes, 0 = no). \\
			
			\textbf{Explanatory Variables} & \\
			\textmd{Gender} & 1 = Male, 0 = Female. \\
			\textmd{Age Cohort} & Respondents age categorizes: 1 = 1995-1983, 2 = 1982-1971, 3 = 1970-1959, 4 = 1958-1947, 5 = 1946-1935. \\
			\textmd{Siblings} & Number of siblings at age 15. \\
			\textmd{School Type} & Type of secondary school respondent attended (1 = private, 0 = public). \\
			\textmd{Academic Performance} & Self-rated academic ranking in the 9th grade. \\
			\textmd{Prefecture} & Prefecture of residence upon secondary school graduation. \\
			\textmd{Parental Income} & Estimated family income level at age 15, based on possession of 19 consumer durables. \\
			\textmd{Parental Education} & Highest academic qualification of parents. \\	
			\textmd{Parental Occupation} & Highest occupational prestige of the parents based on the job description, recoded by the EGP class scheme. \\	
			\bottomrule
		\end{tabularx}
	\end{table}

	\subsection{Descriptive Statistics}
	
	%\subsubsection{Variable Balance}
	
\begin{figure}[hbt!]
		\centering
		\includegraphics[width=\textwidth]{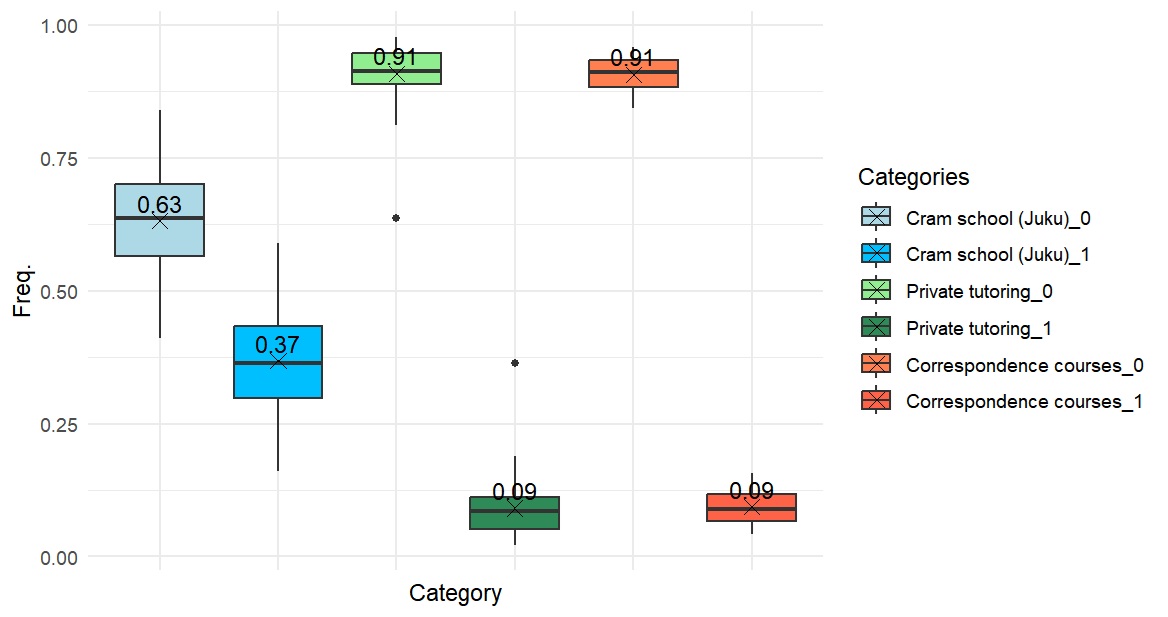}
		\caption{Box plot comparing participation rate of three extracurricular educations from 47 prefectures in Japan.}
		\label{fig6}
	\end{figure}

	\figref{fig6} presents the differences in participation rates across different types of extracurricular education in Japan. Specially, the average participation rate in private tutoring before high school is 9.7\%. However, there is notable variation among prefectures. Tokyo leads with a highest 18.8\% participation rate, while Iwate records the lowest at 2.2\%. In three urban centers, Tokyo's participation rate stands at 18.8\%, Osaka at 16.0\%, and Aichi, home to Nagoya, at 10.4\%. 
	
\begin{figure}[hbt!]
		\centering
		\includegraphics[width=\textwidth]{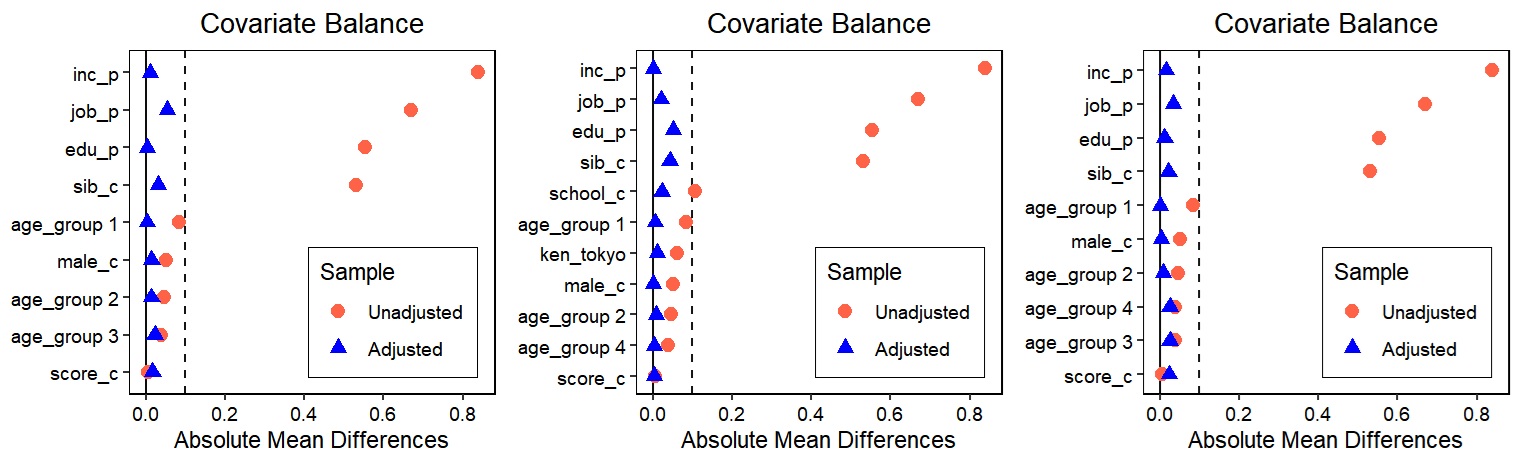}
		\caption{Standardized mean difference of for different outcomes. Left: ITI; middle: ITE; right: ITOP}
		\label{fig0}
	\end{figure}

	% balance check	
	\figref{fig0} presents the balance of covariates between treatment and comparison groups before and after inverse probability weighting. It shows that balance was improved after adjustment, bringing all variables below the threshold of 0.1 for absolute mean differences, indicating good comparability between the treated and control groups.
	
	\tabref{tab2} shows descriptive statistics after multiple imputation \citep{Rubin1987}, a practical method to address missing data in large-scale surveys like the SSM. By replacing null values with plausible estimates based on observed data, this approach reduces bias, preserves variability, and accounts for uncertainty. It ensures robust and reliable analysis, making it essential for accurate results in studies with complex survey data.

	\begin{table}[hbt!]
		\centering
		\caption{Descriptive Statistics}
		\label{tab2}
		\begin{tabularx}{\textwidth}{p{5cm} *{5}{X}}
			\toprule
			\textbf{Variable Name} & \textbf{N} & \textbf{Mean} & \textbf{St. Dev.} & \textbf{Min} & \textbf{Max} \\ 
			\midrule
			\textbf{Outcome Measures} & & & & & \\
			Intergenerational Transmission of Educational Attainment (ITE) & 7,817 & 0.000 & 1.143 & -3.632 & 2.365 \\
			Intergenerational Transmission of Income (ITI) & 7,817 & 0.000 & 1.357 & -4.045 & 5.870 \\
			Intergenerational Transmission of Occupational Prestige (ITOP) & 7,817 & 0.000 & 1.252 & -2.841 & 3.716 \\
			\midrule
			\textbf{Treatment Variables} & & & & & \\
			Cram school (\textit{Juku}) & 7,817 & 0.397 & 0.489 & 0 & 1 \\
			Private tutoring & 7,817 & 0.097 & 0.296 & 0 & 1 \\
			Correspondence courses & 7,817 & 0.101 & 0.301 & 0 & 1 \\
			\midrule
			\textbf{Explanatory Variables} & & & & & \\
			Gender (male) & 7,817 & 0.456 & 0.498 & 0 & 1 \\
			Age  & 7,817 & 52.973 & 16.161 & 20 & 80 \\
			Siblings & 7,817 & 1.832 & 0.915 & 0 & 3 \\
			School Type (Private) & 7,817 & 0.045 & 0.207 & 0 & 1 \\
			Academic Performance & 7,817 & 3.209 & 1.059 & 1 & 5 \\
			Prefecture & 7,817 & 21.471 & 13.013 & 0 & 47 \\
			Parental Income & 7,817 & -0.0002 & 0.950 & -2.230 & 2.342 \\
			Parental Education & 7,817 & 3.014 & 1.040 & 1 & 5 \\
			Parental Occupation & 7,817 & 3.706 & 1.913 & 0 & 7 \\
			\bottomrule
		\end{tabularx}
	\end{table}

	\section{Results}

	\subsection{Average treatment effects}
	%figures	
	%1. hist of CATE			
	\begin{figure}[hbt!]
		\centering
		\includegraphics[width=\textwidth]{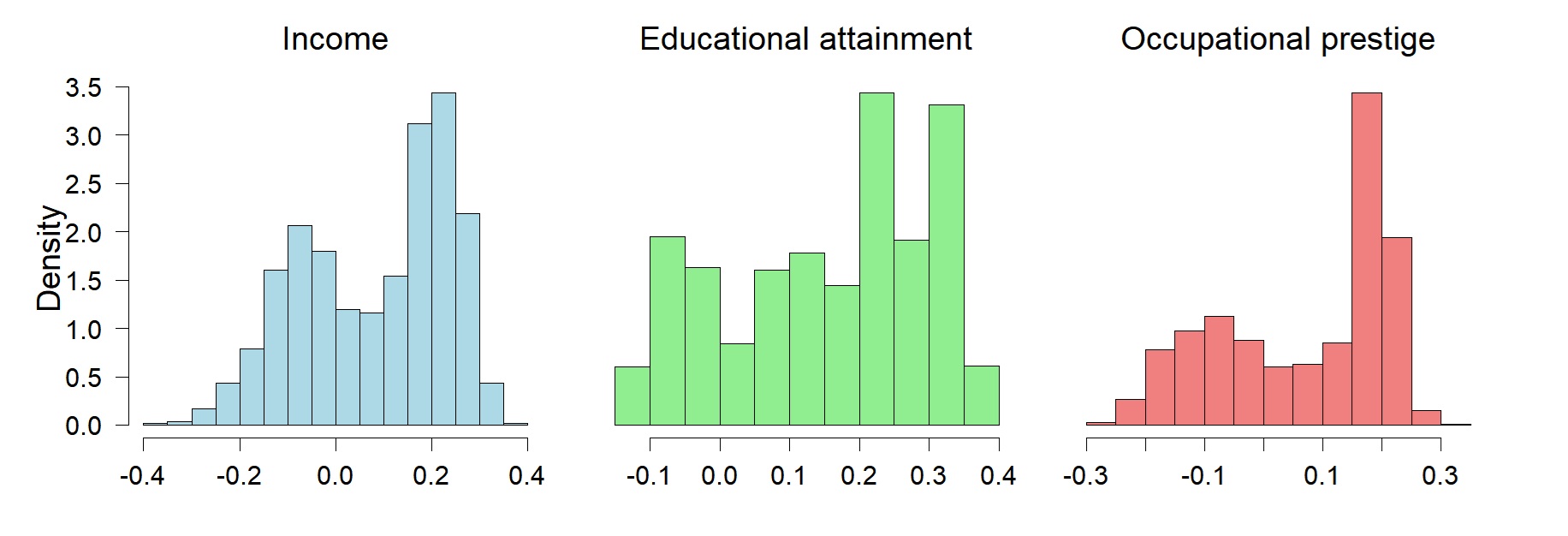}
		\caption{Estimated (conditional) individual level treatment effects for different outcomes. Left: ITI; middle: ITE; right: ITOP}
		\label{fig1}
	\end{figure}
	
\begin{figure}[hbt!]
		\centering
		\includegraphics[width=.7\textwidth]{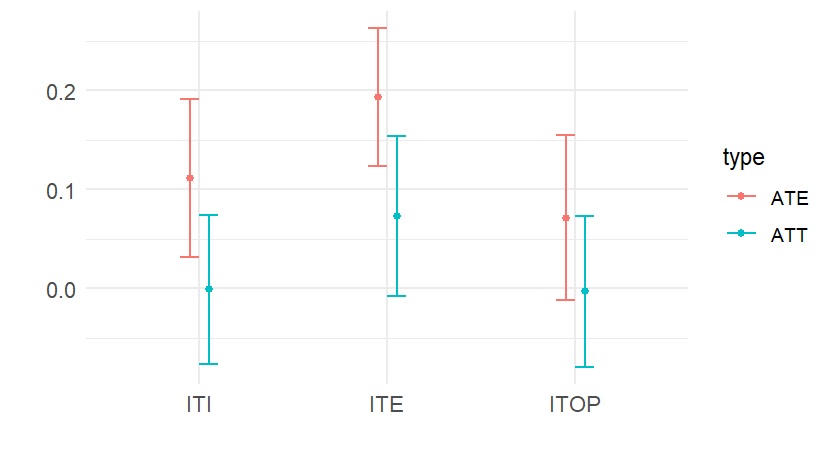}
		\caption{95\% CI of ATE and ATT for the participation of private tutoring in Japan.}
		\label{fig5}
	\end{figure}

	\figref{fig1} presents the estimated treatment effects of private tutoring on income, educational attainment, and occupational prestige. The 95\% confidence intervals for the differences in ATE are as follows: $0.11 \pm 0.08$ for income, $0.19 \pm 0.07$ for educational attainment, and $0.07 \pm 0.08$ for occupational prestige. It suggests that tutoring exerts a positive influence on children's educational attainment and income. However, barely any positive results were detected for the children who received this tutoring, as average treatment effect on the treated (ATT) in \figref{fig5} suggests. A smaller ATT indicates that the treated group may have systematic differences compared to others.

	%2. variable importance(top 5) 
	
\begin{figure}[hbt!]
		\centering
		\includegraphics[width=\textwidth]{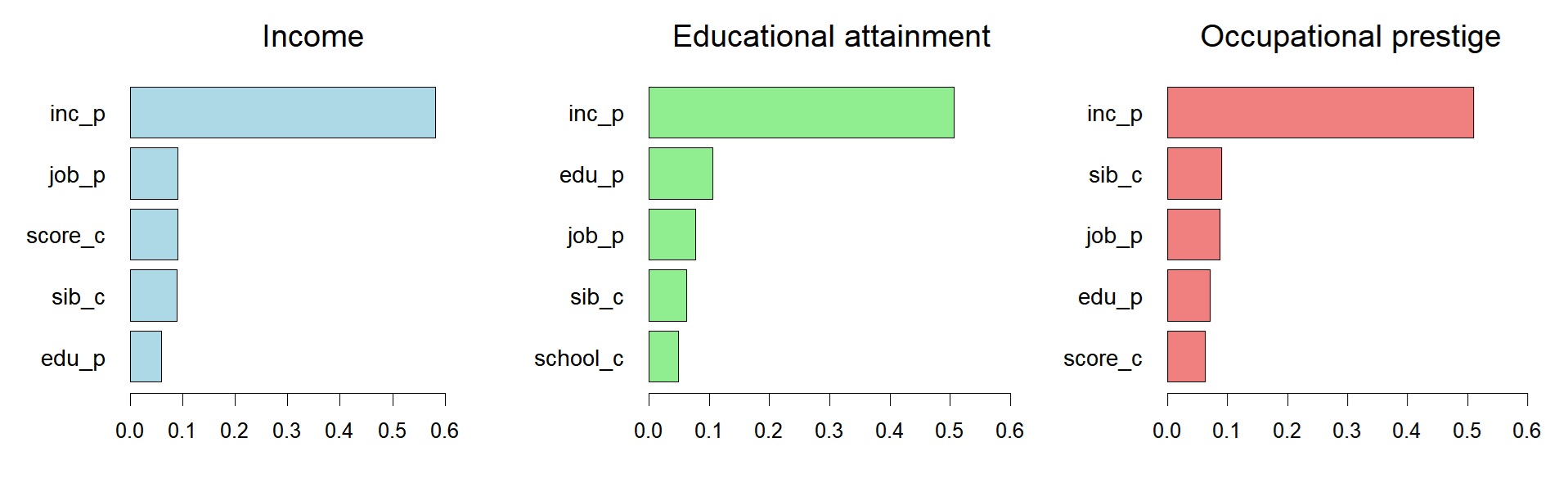}
		\caption{Top 5 variables influencing treatment effect heterogeneity in private tutoring attendance across different outcomes. Left: ITI; middle: ITE; right: ITOP\\
			\textbf{Note:} Variable importance was calculated by the frequency of each variable's use in creating regression tree.}
		\label{fig2}
	\end{figure}

	\figref{fig2} shows the results of variable importance analysis, providing insights into which factors contribute most to the socioeconomic mobility. Among these covariates, parental income (\textit{inc\_p}) emerges as the most influential variable. This finding aligns with existing research emphasizing the critical impact of parental resources on opportunities and outcomes for the next generation \citep{Blanden2005, Chetty2014}. Other variables, including parental education (\textit{edu\_p}), parental occupational (\textit{job\_p}), and the number of siblings (\textit{sib\_c}), also contribute to varying extents. To better understand the heterogeneity in treatment effects, additional analysis can focus on how these variables influence the outcomes differently across subgroups, providing deeper insights into the dynamics of social mobility and inequality.

	\subsection{Heterogenous treatment effects}

\begin{figure}[hbt!]
		\centering
		\includegraphics[width=\textwidth]{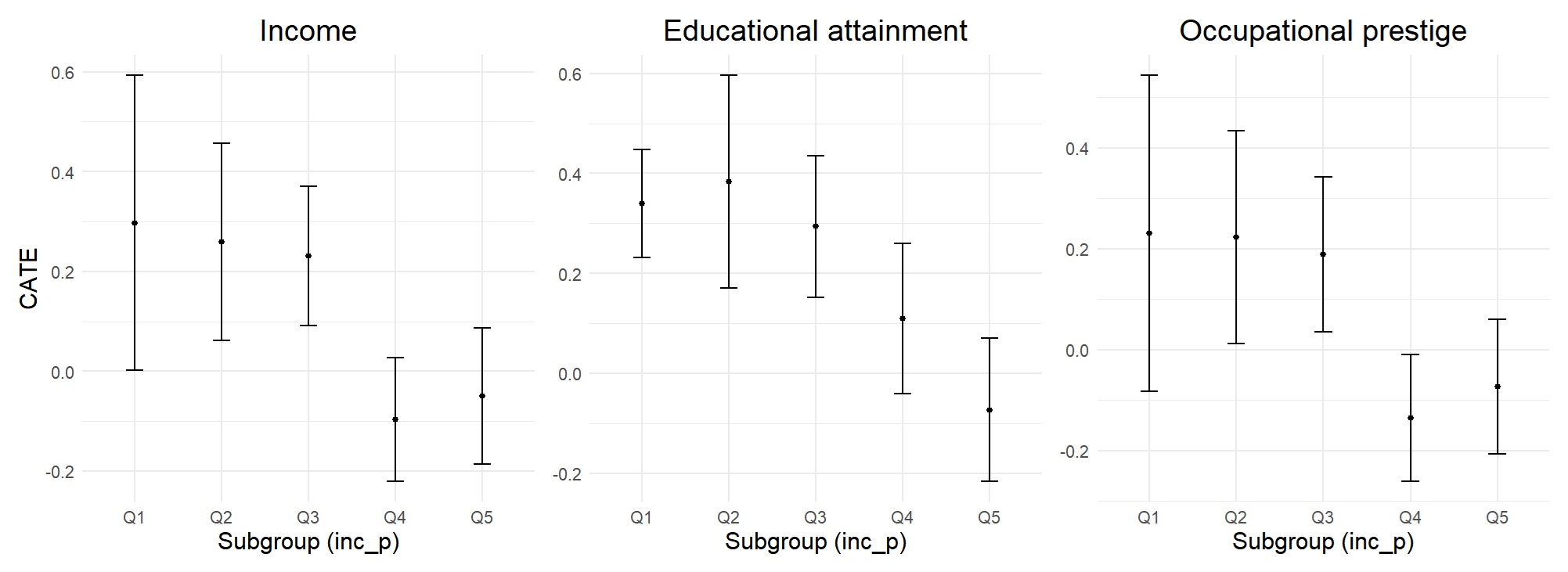}
		\caption{Estimated conditional average treatment effect (CATE) by the most important variable (\( \textit{inc\_p} \)) subgroups. Error bars represent the 95\% confidence intervals.  }
		\label{fig3}
	\end{figure}

	%3. CATE for the most important variable
	\figref{fig3} illustrates the estimated Conditional Average Treatment Effect (CATE) of private tutoring on intergenerational transmission outcomes across quintiles of parental income (\( \textit{inc\_p} \)). For income, the CATE is highest in the lower and middle income quintiles (Q1–Q3), but limits in higher quintiles (Q4–Q5), even showing a potential null or slightly negative effect. A similar pattern is observed in educational attainment and occupational prestige, where the benefits are more significant for lower-income groups and limited effects for wealthier families. This suggests that private tutoring could help bridge socioeconomic gaps for children from disadvantaged backgrounds, highlight the importance of targeting educational interventions to promote equity in social mobility.

	%4. CATE by quantile

\begin{figure}[hbt!]
		\centering
		\includegraphics[width=\textwidth]{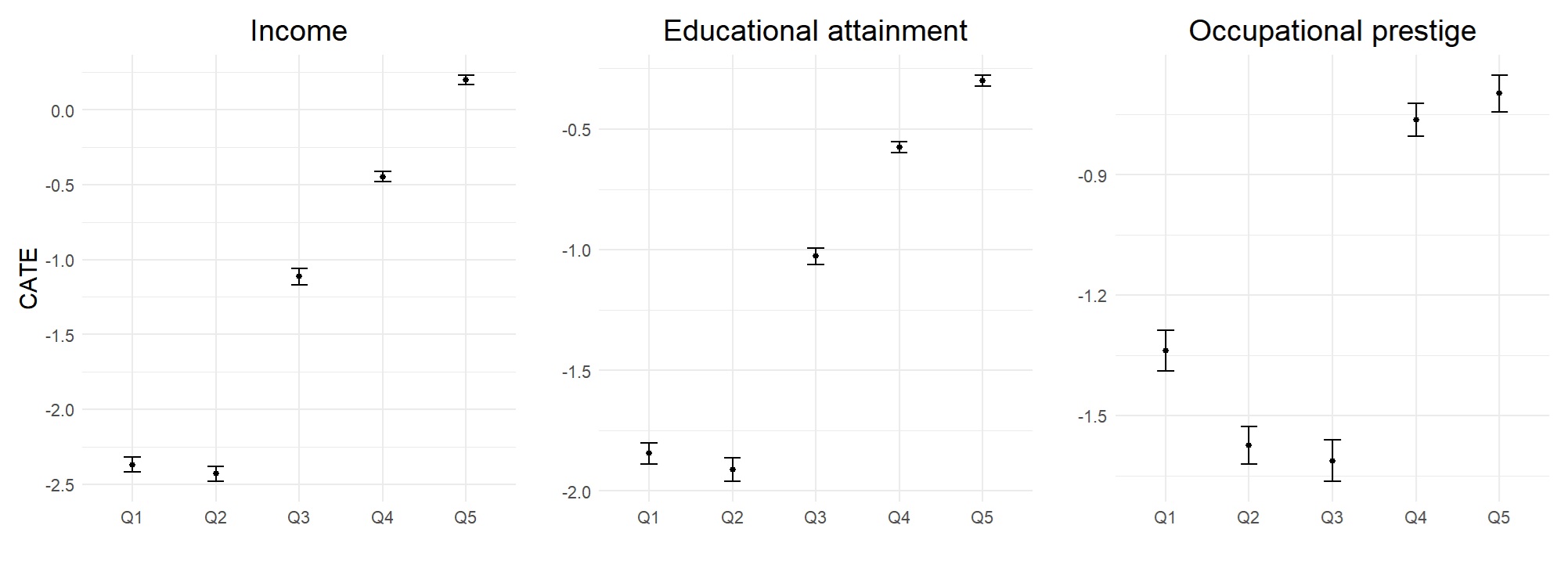}
		\caption{Estimated ATE by quintile ranking for income, educational attainment, and occupational prestige. Error bars represent the 95\% confidence intervals. }
		\label{fig4}
	\end{figure}

	\figref{fig4} shows the estimated ATE of private tutoring on three socioeconomic outcomes across quintiles, ranked by the CATE from the Causal Forest model to highlight distinct subgroups. It reveals a significant negative impact for most subgroups, except for Q5 in income. More negative CATEs appear in lower quintiles, suggesting potential financial strain on lower-income families. This implied the complex effects of private tutoring, emphasizing the need for tailored interventions to maximize benefits.
	
	% the check for test_calibration
	\tabref{tab3} presents the calibration test for Causal Forest, evaluating the model's predictive accuracy and ability to capture heterogeneity. It computes the best linear prediction of the target estimand using the forest prediction (on held-out data) as well as the mean forest prediction as the sole two regressors. Results with strong significance indicate that the Causal Forest models correctly capture heterogeneity in treatment effects across the three socioeconomic outcomes. 
	
	\begin{table}[hbt!]
		\centering
		\caption{Calibration Test for Causal Forest Models.}
		\label{tab3}
		\begin{tabular}{lccc}
			\toprule
			& \multicolumn{3}{c}{\textbf{Dependent variable}} \\
			\cmidrule(lr){2-4}
			& \textbf{Income} & \textbf{Educational attainment} & \textbf{Occupational prestige} \\
			\midrule
			\textmd{mean.forest.prediction} & 0.921* & 1.189*** & 0.900*** \\
			& (0.575) & (0.292) & (0.697) \\
			\textmd{differential.forest.prediction} & 0.803*** & 1.087*** & 0.865*** \\
			& (0.257) & (0.276) & (0.276) \\
			\bottomrule
		\end{tabular}
		%\captionsetup{justification=centering, singlelinecheck=false}
		\caption*{{\textbf{Note:} 	A coefficient of 1 for ``mean.forest.prediction'' suggests correct prediction; whereas a coefficient of 1 for ``differential.forest.prediction'' suggests the forest adequately capture the heterogeneity. *p $<$ 0.1; **p $<$ 0.05; ***p $<$ 0.01. }}
	\end{table}

	%5. Best Linear Prediction
	
	Best linear projection (BLP) could also be applied with Causal Forest\citep{Semenova2021} to assess its ability in capturing treatment effect heterogeneity. By allowing a doubly robust fit to the linear model:  
	
	\begin{equation}
		\hat{\tau}(X_i) = \beta_0 + A_i' \beta_1.
		\label{eq26}
	\end{equation}
	
	\noindent where \textit{$A_i$} can be a subset of the covariate \textit{$X_i$}, it could captures intricate data patterns that traditional linear models might overlook. By regressing observed outcomes on the predicted treatment effects, the coefficient from the BLP test indicates how closely the predictions match the true effects, as \tabref{tab4} suggested.

	\begin{table}[hbt!]
		\centering
		\caption{Best Linear Prediction.}
		\label{tab4}
		\begin{tabularx}{\textwidth}{p{4cm} *{3}{c}}
			\toprule 
			& \multicolumn{3}{c}{\textbf{Dependent variable}} \\ 
			\cmidrule(lr){2-4}
			& \textbf{Income} & \textbf{Educational attainment} & \textbf{Occupational prestige} \\
			%\\[-1.8ex] & (ITI) & (ITE) & (ITOP)\\ 
			\hline \\[-1.8ex] 
			Age group 1 & 0.435$^{**}$ & 0.050 & 0.593$^{*}$ \\ 
			& (0.193) & (0.208) & (0.327) \\ 
			Age group 2 & 0.419$^{**}$ & 0.012 & 0.606$^{*}$ \\ 
			& (0.181) & (0.184) & (0.310) \\ 
			Age group 3 & 0.336$^{**}$ & 0.150 & 0.526$^{*}$ \\ 
			& (0.163) & (0.164) & (0.303) \\ 
			Age group 4 &  & 0.104 & 0.451$^{**}$ \\ 
			&  & (0.134) & (0.176) \\ 
			Sib\_c & $-$0.023 & 0.097$^{**}$ & $-$0.045 \\ 
			& (0.046) & (0.045) & (0.054) \\ 
			Male\_c & 0.115 & 0.156$^{**}$ & 0.073 \\ 
			& (0.081) & (0.072) & (0.083) \\ 
			Score\_c & 0.034 & 0.009 & 0.058 \\ 
			& (0.045) & (0.037) & (0.065) \\ 
			Inc\_p & $-$0.338$^{***}$ & $-$0.107 & $-$0.300$^{**}$ \\ 
			& (0.110) & (0.068) & (0.122) \\ 
			Edu\_p & 0.064 & $-$0.063 & 0.006 \\ 
			& (0.043) & (0.042) & (0.047) \\ 
			Job\_p & $-$0.032 & 0.020 & 0.003 \\ 
			& (0.026) & (0.022) & (0.032) \\ 
			Constant & $-$0.292 & $-$0.025 & $-$0.519 \\ 
			& (0.253) & (0.258) & (0.387) \\ 
			\bottomrule
		\end{tabularx} 
		\caption*{\parbox{\textwidth}{\textbf{Note:} Best linear projection of the conditional average treatment effect; Confidence intervals are cluster- and heteroskedasticity-robust (HC3). Signif. codes: 0 ‘***’; 0.001 ‘**’; 0.01 ‘*’.}}
	\end{table}

	\section{Discussion}
	
	% half page    
	This study suggests that extracurricular education has the potential for positive socioeconomic impacts. Parental income was identified as the most significant factor influencing socioeconomic mobility. In addition, we identified the groups most affected, revealing significant heterogeneity in treatment effects between households. Focusing on the treatment effects, our study contributes to a deeper understanding of how targeted interventions can yield differing outcomes across social groups. 
	
	First, using ML techniques distinguishes our research from traditional approaches to socioeconomic mobility. Previous studies have highlighted the role of education in shaping inter-generational mobility \citep{Becker1979, Jerrim2015}, but most of them relied on the assumption of linearity in estimated models. While ML has been utilized in various econometric applications, its potential in studying socioeconomic mobility is under-explored. By handling high-dimensional data with flexibility, ML offers unique advantages in studying socioeconomic mobility.
	
	Second, this study contributes to the persistent controversy on the results of extracurricular education among different social groups. While some studies support the association between extracurricular activities and positive youth outcomes \citep{Fredricks2006}, others obtained minimal results \citep{Shulruf2010}. Our findings could explain the gap between these different conclusions, indicating that extracurricular education holds potential for enhancing socioeconomic mobility, especially for lower-income families in terms of educational attainment and income. However, the real effects were minimal, as participants from higher-income families, who were more involved, hardly benefited. 
	
	Third, our research indicates that increasing access to extracurricular programs might not be sufficient to close educational and socioeconomic gaps. Instead, it could be more effective to customize strategies for different income groups, such as offering relatively more benefits to those who are disadvantaged \citep{Corak2013}. As we continue to address educational and socioeconomic disparities, the application of ML in policy learning becomes more important for discovering effective policy interventions.

	\section{Conclusion}
	
	This study employs causal machine learning to investigate the average and heterogeneous effects of extracurricular education, with a particular focus on private tutoring. The findings suggest that while shadow education holds the potential for positive socioeconomic impacts, its benefits are often mitigated by the economic disparities among households. This challenges the conventional belief that educational attainment alone guarantees upward mobility, highlighting the importance of addressing customized policy interventions.  
	
	Parental income emerged as the most influential factor in shaping social mobility. By revealing significant heterogeneity in income, educational attainment, and occupational prestige, this study helps to explain the dynamics of social mobility. The use of causal machine learning, particularly Causal Forest, provided a robust framework for analyzing these intricate relationships across a wide range of covariates and potential confounders, offering a more nuanced understanding than traditional methods. 
	
	Several issues remain for further investigation. For example, the factors influencing the demand of extracurricular education requirements, including extended family members, culture, and tradition, remain unclear. Beyond identifying family income as a critical determinant, it is essential to explain the mechanisms driving these disparities, such as social networks, marginal gains, or participant motivation, to better understand how impacts vary across income brackets. Also, the economic efficiency of investing in such activity is yet to be fully explored. Addressing these questions will provide valuable insights for families and policymakers, enabling more informed decisions about the role of extracurricular education in promoting equitable social mobility.

	\subsubsection*{Acknowledgment}
	
	I am deeply grateful to all who supported me in completing this thesis. My heartfelt thanks go to my supervisor, Prof. Oleksandr Movshuk, for his invaluable guidance and encouragement. I extend my gratitude to committee members, Prof. Karato Koji and Prof. Homma Tetsushi, for their constructive feedback. Special thanks to Prof. Ryan T. Moore for his insightful comments and to the 2015 SSM Survey Management Committee for the data access. I also wish to express my sincere appreciation to the Asahi International Education Foundation for their scholarship support. Finally, I am profoundly grateful to my family and friends for their unwavering love and support. Thank you all for being an integral part of this journey.
	
	\subsubsection*{Availability of data and material}
	The data for this analysis was provided by the Social Science Japan Data Archive, Center for Social Research and Data Archives, Institute of Social Science, The University of Tokyo. The code used to construct the analytical dataset can be shared upon request. 
	
	\subsubsection*{Code availability}
	The code used to for the statistical analysis can be found in the Appendix.

	\clearpage
	\subsubsection*{Appendix: Software code}
	
	%\subsubsection*{Causal Forest Estimations}
	\begin{lstlisting}
### Causal Forest Estimations ###
## 1. Install packages / load libraries  ##
install.packages("grf")
library(grf)

## 2. Run Causal Forest models ##
df2015$ken_c.exp <- model.matrix(~ factor(df2015$ken_15) + 0) # To avoid dummy variable trap
df2015$age_group <- as.numeric(cut(df2015$age_c, breaks = 5, labels = c("1", "2", "3", "4", "5"))) # Create age cohort
X <- model.matrix(~ 0 + ken_c.exp + factor(age_group) + sib_c + male_c + school_c + score_c + inc_p + edu_p + job_p, data = df2015)
cf_weights <- df2015$weight # Post-stratification weights
cf_1 <- causal_forest(X, 
		      Y, #the outcome of interest
		      W, #W is the binary treatment variable
		      tune.parameters = "all", 
		      seed = 123, 
		      sample.weights = cf_weights)

## 3. Select variables  ##
varimp <- variable_importance(cf_1)
selected.idx <- which(varimp > mean(varimp)) # Mean selection

## 4. Predicted outcome for the selected variables only  ##
cf_pred_1 <- rep(NA, length(Y))
cf_1.1 <- causal_forest(X[,selected.idx], 
		        Y, 
		        W, 
		        tune.parameters = "all", 
		        seed = 123, 
		        sample.weights = cf_weights)
cf_pred_1 <- predict(cf_1.1, X[,selected.idx])$predictions

## 5. Estimate treatment effects  ##
ATE_cf_1 <- average_treatment_effect(cf_1.1, target.sample = "all")
print(paste("95% of CI for difference in ATE: ", round(ATE_cf_1[1],2), "+-", round(ATE_cf_1[2]*1.96,2)))
ATT_cf_1 <- average_treatment_effect(cf_1.1, target.sample = "treated")
print(paste("95% of CI for difference in ATT: ", round(ATT_cf_1[1],2), "+-", round(ATT_cf_1[2]*1.96,2)))

## 5.1 Subgroup analysis ##
## parental income (inc_p) as example
df2015$inc_p_group <- cut(df2015$inc_p, 
		          breaks = quantile(df2015$inc_p, probs = seq(0, 1, 0.2), na.rm = TRUE), include.lowest = TRUE)
subgroups <- levels(df2015$inc_p_group)
results <- data.frame(subgroup = integer(),
		      CATE = numeric(),
		      CI_lower = numeric(),
		      CI_upper = numeric())
for (i in subgroups) {
  subset <- which(df2015$inc_p_group == i)
  cate_sub <- average_treatment_effect(cf_1.1, subset = subset)
  cate <- cate_sub[1]
  CI <- 1.96 * cate_sub[2]
  results <- rbind(results, data.frame(subgroup = i, 
  CATE = cate, 
  CI_lower = cate - CI,
  CI_upper = cate + CI))}
results$subgroup <- factor(results$subgroup, 
			   levels = levels(df2015$inc_p_group))

## 5.2 Data-driven subgroups ##
## ITI as example

## Concatenate the results.
n.folds <- 5
indices <- split(seq(n), sort(seq(n) %% n.folds))
num.rankings <- 5

tau.hat <- cf_pred_1
e.hat <- cf_1.1$W.hat # P[W=1|X]
m.hat <- cf_1.1$Y.hat # E[Y|X]
res_iti <- lapply(indices, function(idx) {
	# Estimating mu.hat(X, 1) and mu.hat(X, 0) in held-out sample
	mu.hat.0 <- m.hat - e.hat * tau.hat 
	mu.hat.1 <- m.hat + (1 - e.hat) * tau.hat 
	# AIPW scores
	aipw.scores <- (tau.hat	
			+ W[idx] / e.hat * (Y[idx] - mu.hat.1)
			- (1-W[idx]) / (1-e.hat) * (Y[idx]-mu.hat.0))
	# Rank observations on held-out sample based on estimated CATE.
	tau.hat.quantiles <- quantile(tau.hat, 
			probs = seq(0, 1, length.out = num.rankings+1))
	ranking <- cut(tau.hat, tau.hat.quantiles, 
		       include.lowest = TRUE, 
		       labels = paste0("Q", seq(num.rankings)))
	# Store results
	data.frame(aipw.scores, tau.hat, ranking=factor(ranking), outcome=Y[idx], treatment=W[idx])
	})
res_iti <- do.call(rbind, res_iti)

## Average AIPW scores for the treatment effect within each ranking
ate_iti <- lm(aipw.scores ~ 0 + ranking, data=res_iti)
ate_iti <- coeftest(ate_iti, vcov=vcovHC(ate_iti, type='HC2'))
summary(ate_iti)

ate_iti <- data.frame("aipw", paste0("Q", seq(num.rankings)), ate_iti[,1:2])
colnames(ate_iti) <- c("method", "ranking", "estimate", "std.err")
rownames(ate_iti) <- NULL
tapply(res_iti$aipw.scores, res_iti$ranking, mean)
print(ate_iti)

## 6. Best linear projection ## 
best_linear_projection(cf_1.1, X[,selected.idx])
test_calibration(cf_1.1)
	\end{lstlisting}

	\clearpage
	\bibliographystyle{apalike}
	\bibliography{references}

\begin{thebibliography}{}

\bibitem[Abdullah et~al., 2013]{Abdullah.etal.2013}
Abdullah, A., Doucouliagos, H., and Manning, E. (2013).
\newblock Does education reduce income inequality? a meta-regression analysis.
\newblock {\em Journal of Economic Surveys}.
\newblock First published: 19 December 2013.

\bibitem[Athey and Imbens, 2016]{Athey2016}
Athey, S. and Imbens, G. (2016).
\newblock Recursive partitioning for heterogeneous causal effects.
\newblock {\em Proceedings of the National Academy of Sciences},
  113(27):7353--7360.

\bibitem[Athey et~al., 2019]{Athey2019grf}
Athey, S., Tibshirani, J., and Wager, S. (2019).
\newblock {Generalized random forests}.
\newblock {\em The Annals of Statistics}, 47(2):1148 -- 1178.

\bibitem[Athey and Wager, 2019]{Athey2019}
Athey, S. and Wager, S. (2019).
\newblock Estimating treatment effects with causal forests: An application.

\bibitem[Austin, 2011]{Austin2011}
Austin, P.~C. (2011).
\newblock An introduction to propensity score methods for reducing the effects
  of confounding in observational studies.
\newblock {\em Multivariate Behavioral Research}, 46(3):399--424.
\newblock Published online 2011 Jun 8.

\bibitem[Becker and Tomes, 1979]{Becker1979}
Becker, G.~S. and Tomes, N. (1979).
\newblock An equilibrium theory of the distribution of income and
  intergenerational mobility.
\newblock {\em Journal of Political Economy}, 87(6):1153--1189.

\bibitem[Belloni et~al., 2014]{Belloni2014}
Belloni, A., Chernozhukov, V., and Hansen, C. (2014).
\newblock High-dimensional methods and inference on structural and treatment
  effects.
\newblock {\em Journal of Economic Perspectives}, 28(2):29--50.

\bibitem[Black and Devereux, 2011]{Black2011}
Black, S.~E. and Devereux, P.~J. (2011).
\newblock Recent developments in intergenerational mobility.
\newblock {\em Handbook of labor economics}, 4:1487--1541.

\bibitem[Blanden et~al., 2005]{Blanden2005}
Blanden, J., Gregg, P., and Machin, S. (2005).
\newblock Intergenerational mobility in europe and north america.
\newblock {\em Report supported by the Sutton Trust, Centre for Economic
  Performance, London School of Economics}.

\bibitem[Bray, 2007]{Bray2007}
Bray, M. (2007).
\newblock {\em The Shadow Education System: Private Tutoring and Its
  Implications for Planners}.
\newblock Fundamentals of educational planning, 61. UNESCO IIEP, Paris.

\bibitem[Buchmann et~al., 2010]{Buchmann.etal.2010}
Buchmann, C., Condron, D.~J., and Roscigno, V.~J. (2010).
\newblock {Shadow Education, American Style: Test Preparation, the SAT and
  College Enrollment}.
\newblock {\em Social Forces}, 89(2):435--461.

\bibitem[Chernozhukov et~al., 2018]{Chernozhukov2018}
Chernozhukov, V., Chetverikov, D., Demirer, M., Duflo, E., Hansen, C., Newey,
  W., and Robins, J. (2018).
\newblock {Double/debiased machine learning for treatment and structural
  parameters}.
\newblock {\em The Econometrics Journal}, 21(1):C1--C68.

\bibitem[Chetty et~al., 2014]{Chetty2014}
Chetty, R., Hendren, N., Kline, P., and Saez, E. (2014).
\newblock Where is the land of opportunity? the geography of intergenerational
  mobility in the united states.
\newblock {\em The quarterly journal of economics}, 129(4):1553--1623.

\bibitem[Cockx et~al., 2023]{Cockx2023}
Cockx, B., Lechner, M., and Bollens, J. (2023).
\newblock Priority to unemployed immigrants? a causal machine learning
  evaluation of training in belgium.
\newblock {\em Labour Economics}, 80:102306.

\bibitem[Coqueret, 2021]{Coqueret2020}
Coqueret, G. (2021).
\newblock Machine learning in finance: From theory to practice.
\newblock {\em Quantitative Finance}, 21(1):9--10.

\bibitem[Corak, 2013]{Corak2013}
Corak, M. (2013).
\newblock Income inequality, equality of opportunity, and intergenerational
  mobility.
\newblock {\em Journal of Economic Perspectives}, 27(3):79--102.

\bibitem[Entrich, 2015]{Entrich2015}
Entrich, S.~R. (2015).
\newblock The decision for shadow education in japan: Students’ choice or
  parents’ pressure?
\newblock {\em Social Science Japan Journal}, 18(2):193--216.

\bibitem[Filmer and Pritchett, 2001]{Filmer2001}
Filmer, D. and Pritchett, L.~H. (2001).
\newblock Estimating wealth effects without expenditure data---or tears: An
  application to educational enrollments in states of india.
\newblock {\em Demography}, 38(1):115--132.

\bibitem[Fredricks and Eccles, 2006]{Fredricks2006}
Fredricks, J.~A. and Eccles, J.~S. (2006).
\newblock Is extracurricular participation associated with beneficial outcomes?
  concurrent and longitudinal relations.
\newblock {\em Developmental psychology}, 42(4):698.

\bibitem[Fukai et~al., 2021]{Fukai2021}
Fukai, T., Ichimura, H., and Kawata, K. (2021).
\newblock Describing the impacts of covid-19 on the labor market in japan until
  june 2020.
\newblock {\em The Japanese Economic Review}, 72(3):439--470.

\bibitem[Gregorio and Lee, 2002]{Gregorio&Lee2002}
Gregorio, J.~D. and Lee, J.-W. (2002).
\newblock Education and income inequality: New evidence from cross-country
  data.
\newblock {\em Review of Income and Wealth}, 48(3):395--416.

\bibitem[Holland, 1986]{Holland1986}
Holland, P.~W. (1986).
\newblock Statistics and causal inference.
\newblock {\em Journal of the American Statistical Association},
  81(396):945--960.

\bibitem[HU et~al., 2016]{HU_FAN_DING_2016}
HU, Y., FAN, W., and DING, W. (2016).
\newblock Does shadow education aggravate inequality of educational outcomes.
\newblock {\em The Eurasia Proceedings of Educational and Social Sciences},
  4:11–32.

\bibitem[Imbens and Rubin, 2015]{Imbens&Robin2015}
Imbens, G.~W. and Rubin, D.~B. (2015).
\newblock {\em {Causal Inference for Statistics, Social, and Biomedical
  Sciences}}.
\newblock Number 9780521885881 in Cambridge Books. Cambridge University Press.

\bibitem[Jerrim and Macmillan, 2015]{Jerrim2015}
Jerrim, J. and Macmillan, L. (2015).
\newblock Income inequality, intergenerational mobility, and the great gatsby
  curve: Is education the key?
\newblock {\em Social Forces}, 94(2):505--533.

\bibitem[Kang and Schafer, 2007]{Joseph2007}
Kang, J. D.~Y. and Schafer, J.~L. (2007).
\newblock {Demystifying Double Robustness: A Comparison of Alternative
  Strategies for Estimating a Population Mean from Incomplete Data}.
\newblock {\em Statistical Science}, 22(4):523 -- 539.

\bibitem[Kanomata et~al., 2008]{Kanomata2008}
Kanomata, N., Tanabe, S., and Takenoshita, H. (2008).
\newblock Ssm occupational code and international measures of occupational
  status: Conversion into egp class schema, siops and isei.
\newblock {\em Problems in Measurement and Analysis in Social Surveys (2005 SSM
  Survey Series 12)}, pages 69--94.

\bibitem[Masahiro and Wataru, 2023]{Narisawa2023}
Masahiro, N. and Wataru, Y. (2023).
\newblock Two types of class attainment mediated by education: The effects of
  high school rank and type on class destination (in japanese).
\newblock {\em Japanese Sociological Review}, 74(1):34--50.

\bibitem[Ministry~of Education and Technology, 2008]{MEXT2008}
Ministry~of Education, Culture, S.~S. and Technology (2008).
\newblock {\em Survey Concerning Out-of-School Learning Activities of
  Schoolchildren (In Japanese)}.
\newblock MEXT.

\bibitem[Mitchell, 1997]{Mitchell1997}
Mitchell, T.~M. (1997).
\newblock {\em Machine Learning}.
\newblock McGraw-Hill.

\bibitem[Nakamura et~al., 2023]{Nakamura2023}
Nakamura, K., Kaneda, T., and Tanaka, H. (2023).
\newblock Inequality of extracurricular educational opportunities: Measuring
  the inequality of paid learning opportunities using the kakwani coefficient
  (in japanese).
\newblock Working Paper 350, School of Economics, University of Toyama.
\newblock Technical Report, No.350, 2023.01, School of Economics, University of
  Toyama.

\bibitem[Nakano, 1973]{Nakano1973}
Nakano, H. (1973).
\newblock Ministry of education "one hundred year history of the education
  system".
\newblock {\em The Japanese Journal of Educational Research}, 40(1):52--54.

\bibitem[Nakazawa, 2010]{Nakazawa2010}
Nakazawa, W. (2010).
\newblock Latent class analysis of intergenerational educational attainment (in
  japanese).
\newblock {\em Japanese Sociological Review}, 61(2):112--129.

\bibitem[Robins et~al., 1994]{Robins1994}
Robins, J.~M., Rotnitzky, A., and Zhao, L.~P. (1994).
\newblock Estimation of regression coefficients when some regressors are not
  always observed.
\newblock {\em Journal of the American statistical Association},
  89(427):846--866.

\bibitem[Rosenbaum, 1987]{Rosenbaum1987}
Rosenbaum, P.~R. (1987).
\newblock Model-based direct adjustment.
\newblock {\em Journal of the American statistical Association},
  82(398):387--394.

\bibitem[Rosenbaum and Rubin, 1983]{Rosenbaum1983}
Rosenbaum, P.~R. and Rubin, D.~B. (1983).
\newblock The central role of the propensity score in observational studies for
  causal effects.
\newblock {\em Biometrika}, 70(1):41--55.

\bibitem[Rubin, 1979]{Rubin1979}
Rubin, D.~B. (1979).
\newblock Using multivariate matched sampling and regression adjustment to
  control bias in observational studies.
\newblock {\em Journal of the American Statistical Association},
  74(366a):318--328.

\bibitem[Rubin, 1987]{Rubin1987}
Rubin, D.~B. (1987).
\newblock {\em Multiple Imputation for Nonresponse in Surveys}.
\newblock Wiley Series in Probability and Statistics. John Wiley \& Sons, Inc.,
  New York.

\bibitem[Ryabov, 2020]{Ryabov2020}
Ryabov, I. (2020).
\newblock Intergenerational transmission of socio-economic status: The role of
  neighborhood effects.
\newblock {\em Journal of Adolescence}, 80:84--97.

\bibitem[Semenova and Chernozhukov, 2020]{Semenova2021}
Semenova, V. and Chernozhukov, V. (2020).
\newblock Debiased machine learning of conditional average treatment effects
  and other causal functions.
\newblock {\em The Econometrics Journal}, 24(2):264--289.

\bibitem[Shulruf, 2010]{Shulruf2010}
Shulruf, B. (2010).
\newblock Do extra-curricular activities in schools improve educational
  outcomes? a critical review and meta-analysis of the literature.
\newblock {\em International Review of Education}, 56:591--612.

\bibitem[Thompson and Simmons, 2013]{Thompson&Simmons2013}
Thompson, R. and Simmons, R. (2013).
\newblock Social mobility and post-compulsory education: revisiting boudon’s
  model of social opportunity.
\newblock {\em British Journal of Sociology of Education}, 34(5-6):744--765.

\bibitem[Tibshirani et~al., 2024]{GRF2024}
Tibshirani, J., Athey, S., Sverdrup, E., and Wager, S. (2024).
\newblock {\em grf: Generalized Random Forests}.
\newblock R package version 2.3.2.

\end{thebibliography}

\end{document}